\documentclass[10pt,conference]{IEEEtran}
\IEEEoverridecommandlockouts

\usepackage{cite}
\usepackage{amsmath,amssymb,amsfonts}
\usepackage{graphicx}
\usepackage{textcomp}
\usepackage{xcolor}
\usepackage{booktabs}
\usepackage{multirow}
\usepackage{url}
\usepackage{enumitem}
\usepackage{float}
\setlist{nosep,leftmargin=*}
\graphicspath{{figures/}}

\begin{document}

\title{Where Atom Loss Lands Matters: Decoder-Aware\\
Risk Deposition in Neutral-Atom QEC}

\author{\IEEEauthorblockN{Xinyi Li}
\IEEEauthorblockA{\textit{Stevens Institute of Technology}\\
Hoboken, USA \\
xli215@stevens.edu}
\and
\IEEEauthorblockN{Yifeng Peng}
\IEEEauthorblockA{\textit{Stevens Institute of Technology}\\
Hoboken, USA \\
ypeng21@stevens.edu}

\and
\IEEEauthorblockN{Ying Wang}
\IEEEauthorblockA{\textit{Stevens Institute of Technology}\\
Hoboken, USA \\
ywang6@stevens.edu}
}

\maketitle

\begin{abstract}
Neutral-atom arrays are emerging as a leading platform for scalable quantum error correction (QEC). Qubits are routed and reused across the array, while detected loss is reported to the decoder as erasure information. Existing neutral-atom compilers optimize this \emph{movement}, including routing, shuttling, and reuse, and often model loss through scalar exposure costs. Yet total exposure is an incomplete statistic for erasure-corrected QEC. It captures how much loss occurs, but not where it lands on the code, which we call its \emph{deposition}. Under the same expected atom-loss budget, different deposition patterns over a code patch induce substantially different logical error rates (LER). We formalize this as \emph{decoder-aware risk deposition} and present CAST, a compiler-side optimization pass that overlays a code-topology sensitivity map on a role-indexed exposure ledger and minimizes a decoder-weighted harm objective under a comparable-exposure constraint, using only local route, role, and seam-cooling actions. Across surface-code memory, physical-scale architecture models, lattice surgery, and decoder-mismatch checks, CAST lowers LER relative to topology-blind exposure minimization, improving on it in 35 of 48 physical-scale settings and by as much as $5.3\times$ where exposure is heterogeneous and routing has slack. The largest gains occur when high exposure and high decoder sensitivity are initially misaligned, giving CAST room to redirect risk toward lower-impact code roles. CAST shows that decoder-aware atom-loss risk deposition can be optimized as a compiler-side pass in neutral-atom QEC.
\end{abstract}


\section{Introduction}

As quantum workloads scale, growing execution costs and hardware noise
strengthen the need for fault-tolerant quantum computing
(FTQC)~\cite{peng2026titan,peng2026quantum}.
Neutral-atom arrays are a promising platform for scalable quantum error correction (QEC) and FTQC. They expose a distinctive interface between architecture, compilation, and fault tolerance. Qubits can be moved and reused through array-level routing~\cite{wang2024atomique,lin2025reuse, ruan2025powermove}, while atom loss can be detected and reported as erasure information to the decoder~\cite{chow2024circuit,perrin2025quantum}. This combination makes loss not only a physical noise process but also a compiler-visible resource: the compiler influences where atoms spend time, which operations they experience, and which code roles inherit the exposure.

\begin{figure}[!t]
\centering
\includegraphics[width=\linewidth]{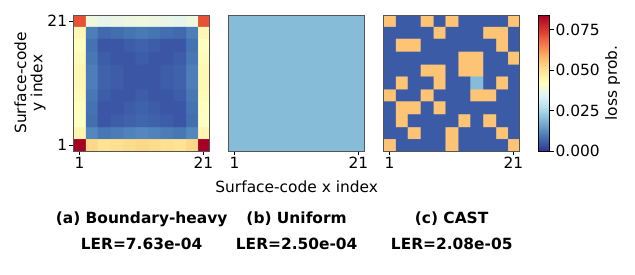}
\caption{\textbf{Equal-budget loss-deposition maps.} Distance-11 surface-code memory patch over 11 syndrome rounds. Each cell is a data-qubit site in surface-code coordinates; color denotes per-round atom-loss probability. All maps have the same expected spacetime loss budget of 26.62 erasures.}
\label{fig:where-loss-lands}
\end{figure}

For erasure-corrected QEC, total exposure is insufficient to characterize decoder-relevant loss risk~\cite{perrin2025quantum}. Two executions with the same expected loss budget can induce different logical error rate (LER) values when loss is deposited on different topological structures. Loss on an interior data-qubit role, a boundary-adjacent role, or a lattice-surgery seam can have different decoding consequences even if the physical exposure is identical. Thus a compiler that minimizes only aggregate movement or aggregate loss can miss a code-level optimization opportunity. This effect is illustrated in Fig.~\ref{fig:where-loss-lands} by a controlled surface-code memory experiment comparing distinct deposition patterns: the expected loss budget is held fixed, but the induced LER changes with the deposition topology.

Prior work has largely developed neutral-atom movement compilation and erasure decoding along separate lines. Neutral-atom compilers optimize routing, shuttling, reuse, and zone traffic, while fault-tolerance work models loss as erasure information for decoding~\cite{wang2024atomique, lin2025reuse,ruan2025powermove,chow2024circuit,perrin2025quantum}. These approaches largely treat compilation and decoding as adjacent stages: the compiler produces a trace, and the decoder handles the resulting faults. What remains underexplored is a compiler objective that uses decoder sensitivity under comparable physical exposure.

CAST is a compiler-side optimization pass for decoder-aware atom-loss risk deposition. It converts a neutral-atom execution trace into a role-indexed exposure ledger, overlays a code-topology sensitivity map, and searches local route, role, and seam-shaping actions that reduce decoder-weighted harm while preserving comparable physical exposure. The key insight is to use decoder sensitivity as a compiler-side ordering of where atom-loss exposure is most consequential. This turns unavoidable loss risk into a placement problem: under a comparable-exposure constraint, CAST steers exposure away from high-consequence code regions and toward lower-impact roles whenever valid local alternatives exist.

\begin{figure*}[!t]
\centering
\includegraphics[width=0.95\linewidth]{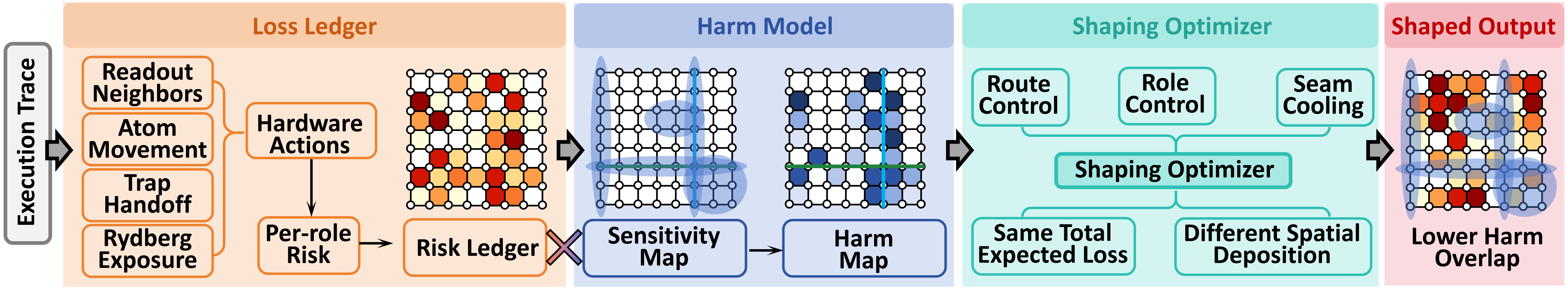}
\caption{\textbf{Methodology overview.} CAST builds a loss ledger from a neutral-atom trace, combines it with a code-sensitivity harm model, and uses local shaping actions to reduce decoder-weighted harm under comparable exposure.}
\label{fig:methodology}
\end{figure*}

Our contributions are summarized as follows:
\begin{itemize}
    \item We identify atom-loss deposition as a compiler-visible fault-tolerance variable: at fixed expected exposure, where loss risk lands on the code can change logical failure.
    \item We formulate decoder-aware deposition as a constrained trace-level objective that minimizes decoder-weighted harm under comparable physical exposure.
    \item We design CAST, a compiler-side optimization pass that combines trace exposure ledgers, code-topology sensitivity maps, and local route, role, and seam-shaping actions.
    \item We evaluate CAST across memory, lattice-surgery, physical-scale, ablation, hardware-sensitivity, and decoder-mismatch studies, showing setting-dependent gains and identifying risk heterogeneity and routing slack as sources of headroom.
\end{itemize}

\section{System Model and Method}


CAST is a trace-level compiler pass from execution traces to decoder-aware placement decisions, as summarized in Fig.~\ref{fig:methodology}. CAST takes a neutral-atom execution trace and a QEC code description as input: the trace specifies how code roles are bound to mobile atoms and scheduled through movement, interaction, readout, and storage. The pass first builds a role-indexed loss ledger, then weights the ledger by code sensitivity, and finally searches local route, role-binding, and seam transformations under a comparable-exposure constraint. The following subsections define these three blocks and the resulting optimization problem.

\subsection{Loss Ledger: Trace to Exposure}

Let $r\in V$ denote a code role, such as a data qubit, check qubit, seam role, or boundary-adjacent role, and let $t$ index execution windows. A neutral-atom trace provides nonnegative exposure counts for movement ($M$), handoff or transport ($H$), entangling gates ($G$), readout/loss detection ($Q$), and idle residence ($I$). CAST converts these counts into per-window loss exposure:
\begin{equation}
R_{r,t}=\alpha_m M_{r,t}+\alpha_h H_{r,t}+\alpha_g G_{r,t}
       +\alpha_q Q_{r,t}+\alpha_i I_{r,t}.
\label{eq:exposure}
\end{equation}
The accumulated role exposure is $R_r=\sum_t R_{r,t}$, and the ledger is the vector $R=\{R_r\}_{r\in V}$. Equal-budget settings normalize $\sum_r R_r$ to a fixed budget; physical-scale settings induce $R$ from architecture parameters. The ledger records where physical exposure lands, while the harm model supplies decoder consequence.

\subsection{Harm Model: Code Sensitivity}

CAST assigns each code role $r$ a normalized sensitivity $S_r$. The map is constructed from code geometry: memory roles combine sampled logical supports, check-neighborhood activity, boundary membership, and a central seam proxy, while lattice-surgery roles additionally include paired merge-boundary supports that enter seam observables. Higher $S_r$ means that depositing exposure on role $r$ contributes more to placement harm. CAST ranks valid candidate ledgers with the decoder-weighted harm objective:
\begin{equation}
\mathrm{DWH}(R)=\sum_{r\in V} S_rR_r+\lambda_s\,\mathrm{SeamRisk}(R),
\label{eq:dwh}
\end{equation}
where $\mathrm{SeamRisk}(R)$ is the fraction of exposure deposited on boundary/seam-proxy roles for memory or on merge-seam observable supports for lattice surgery. The coefficient $\lambda_s$ sets the relative weight of this seam-proximal exposure term, so the objective orders candidates by both role sensitivity and seam-localized risk when comparing placements.

\begin{figure}[!t]
\centering
\includegraphics[width=0.9\linewidth]{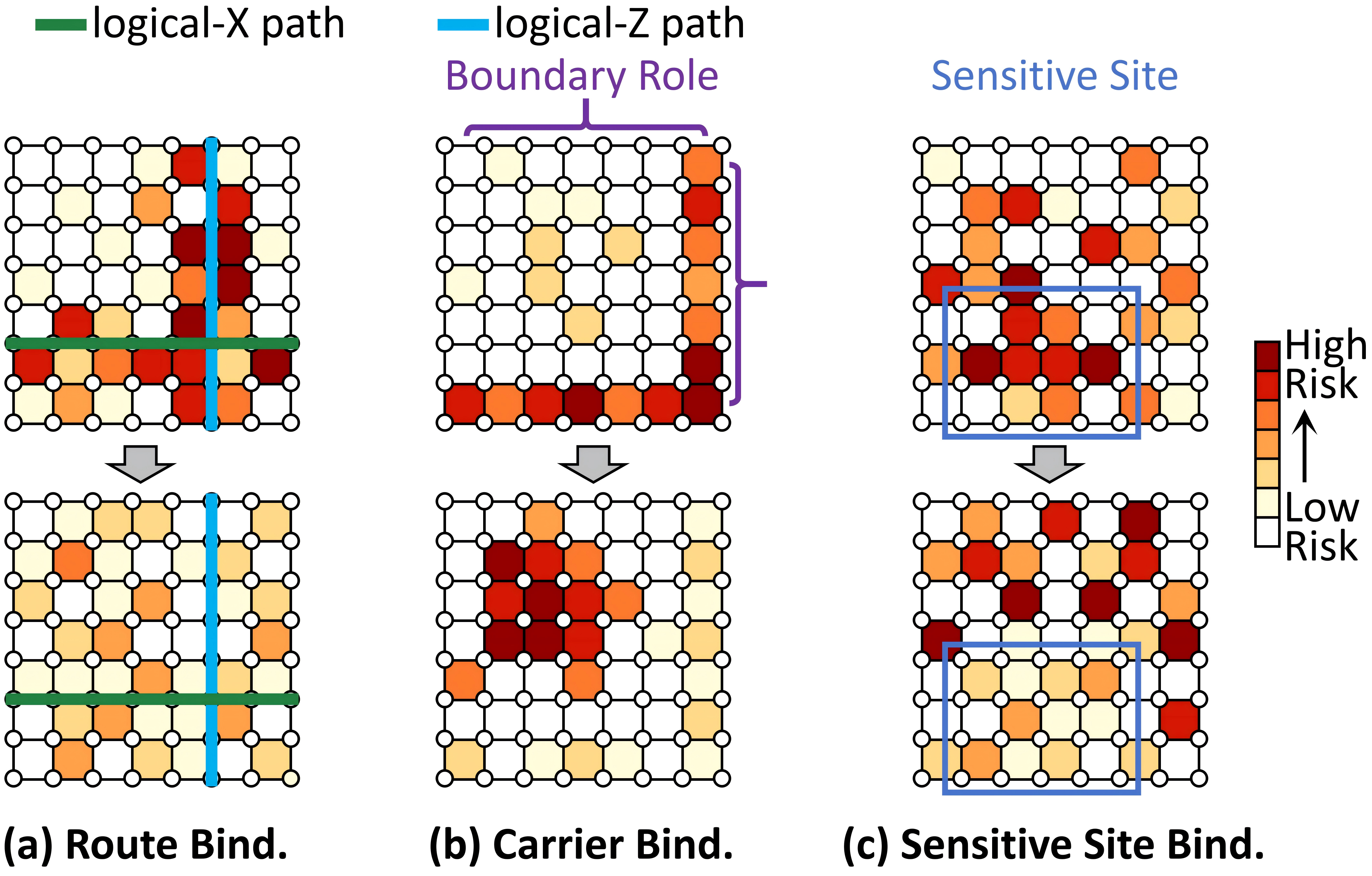}
\caption{\textbf{Local shaping mechanism.} CAST changes which feasible route, role binding, or surgery-seam placement carries exposure.}
\label{fig:local-mechanism}
\end{figure}

\begin{figure*}[!t]
\centering
\includegraphics[width=0.99\textwidth]{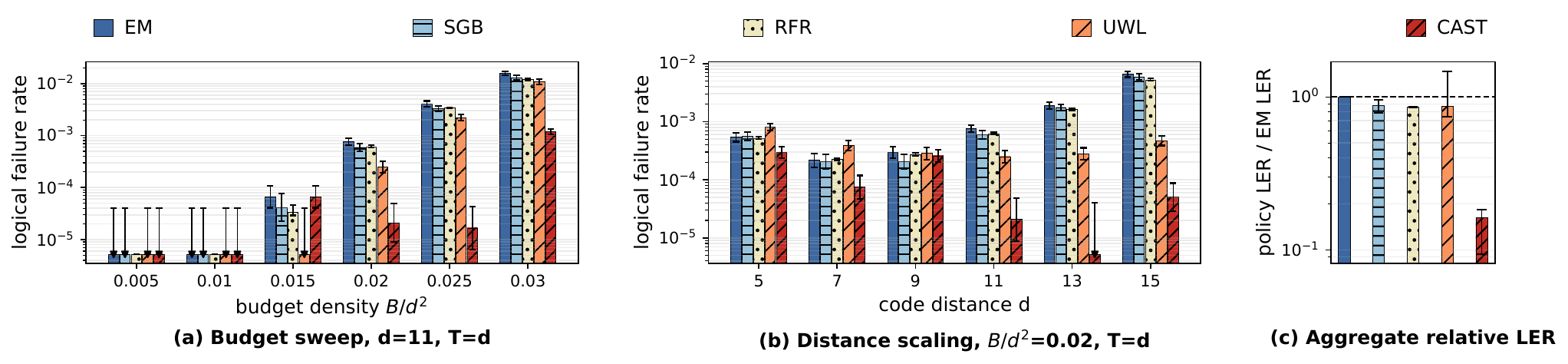}
\caption{\textbf{Equal-budget controlled memory experiment.} Panel (a) sweeps loss budget at $d=11$ and $T=d$; panel (b) sweeps code distance at fixed budget density and $T=d$; panel (c) aggregates measurable paired points as policy/EM LER ratios. Lower LER and lower ratios are better.}
\label{fig:equal-budget-controlled}
\end{figure*}

\subsection{Shaping Actions}

For a trace $T$, $\mathcal{A}(T)$ denotes CAST's local action space: trace-valid rewrites that preserve the circuit, syndrome schedule, logical operation, and comparable total exposure while changing where that exposure lands on code roles. CAST instantiates this space with three action families, as shown in Fig.~\ref{fig:local-mechanism}: route shaping chooses feasible route alternatives with comparable movement cost, role shaping matches an exposure histogram to less sensitive code roles, and seam cooling moves deposition away from boundary/seam neighborhoods in merge-split operations.

\subsection{CAST Optimization Pass}

CAST selects the trace-valid action that minimizes decoder-weighted harm while keeping total exposure comparable:
\begin{equation}
\min_{a\in\mathcal{A}(T)} \mathrm{DWH}(R(a))
\quad
\mathrm{s.t.}\quad
\left|\sum_r R_r(a)-B_T\right|\le \epsilon_B .
\label{eq:problem}
\end{equation}
Here $R(a)$ is the ledger induced by action $a$, $B_T$ is the reference total exposure for trace $T$, and $\epsilon_B$ is the allowed tolerance. The pass constructs Eq.~\eqref{eq:exposure}, scores candidates with Eq.~\eqref{eq:dwh}, and returns the candidate with the lowest decoder-weighted harm satisfying the constraint in Eq.~\eqref{eq:problem}.


\section{Evaluation}

\subsection{Methodology}

\begin{figure}[!t]
\centering
\includegraphics[width=\linewidth]{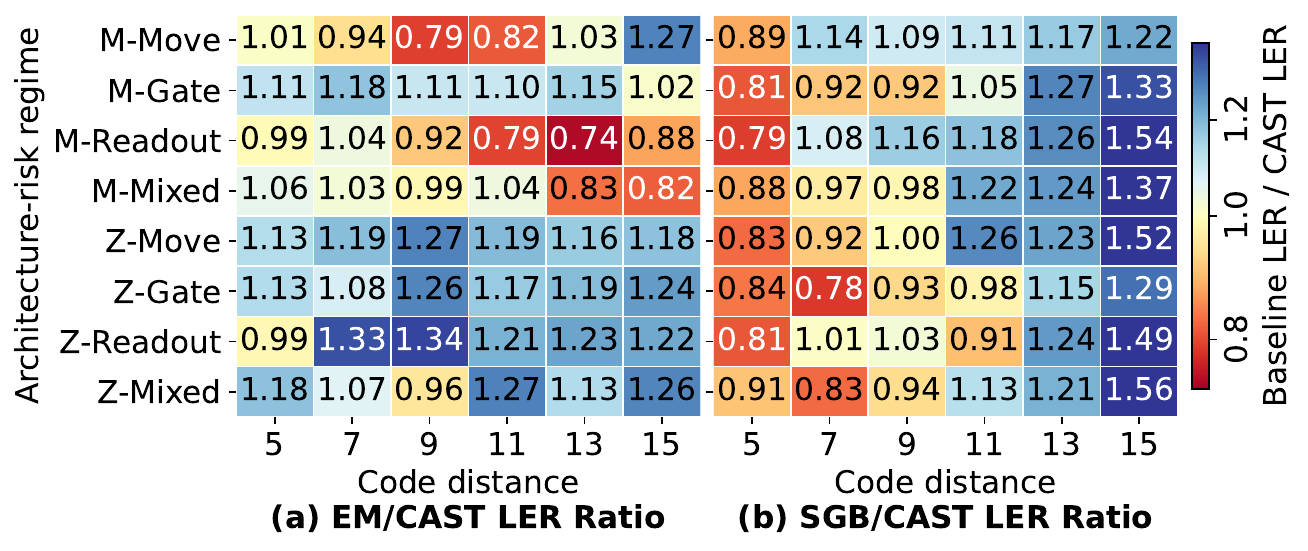}
\caption{\textbf{Calibrated physical-scale comparison.} The two heatmaps report EM/CAST and SGB/CAST LER ratios. Within each heatmap, rows are architecture-risk regimes and columns are code distances. Values above one favor CAST; M/Z denote monolithic/zoned layouts.}
\label{fig:e5-physical-robustness}
\end{figure}

\noindent\textbf{Benchmarks.} We evaluate rotated surface-code memory circuits ($d=5$--$15$, $T=d$) and lattice-surgery merge-split circuits. Traces come from controlled movement-aware routing and monolithic/zoned physical-scale models. Per-role erasure probabilities induce loss samples; we report LER from exact GF(2) erasure decoding and use Stim/PyMatching for representative circuit-level checks. \textbf{Baselines.} We compare CAST against EM (scalar exposure minimization), SGB (fixed boundary/seam guard bands), RFR (random feasible routing), and UWL (idealized equal-budget wear leveling).

\begin{figure}[!t]
\centering
\includegraphics[width=0.8\linewidth]{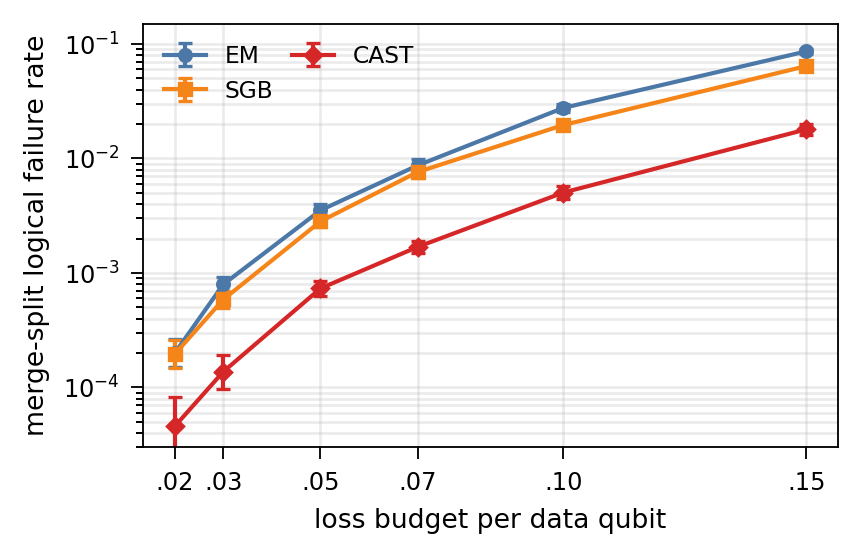}
\caption{\textbf{Lattice-surgery seam-cooling sweep.} Distance-3 equal-budget merge-split experiment under mixed loss sources, comparing EM, SGB, and CAST. Error bars show Wilson intervals; lower is better.}
\label{fig:e6-seam-cooling}
\end{figure}

\subsection{Effectiveness Across Settings}

\textbf{CAST reduces logical failure at the same atom-loss budget by steering exposure away from decoder-sensitive regions.} The equal-budget memory runs in Fig.~\ref{fig:equal-budget-controlled} compare policies under matched expected exposure. CAST has the lowest aggregate ratio among the plotted policies (about $0.14\times$ EM on measurable paired points). Because expected exposure is matched across policies, this improvement isolates loss placement as the causal variable. SGB helps by protecting fixed structures, but its static mask cannot track geometry-dependent sensitivity; UWL remains an idealized non-routing comparison reference.

\textbf{CAST remains effective under calibrated physical-scale exposure.} Exposure is induced by monolithic (M) and zoned neutral-atom (Z) layouts with movement-, gate-, readout-, or mixed-dominated loss sources, as shown in Fig.~\ref{fig:e5-physical-robustness}. Across these settings, CAST improves over EM in 35 of 48 cases and over SGB in 28 of 48 cases, showing that decoder-aware deposition remains useful beyond controlled budgets. Settings where simpler baselines remain competitive mark the boundary condition: CAST gains require risk heterogeneity and routing slack.

\textbf{Seam-aware shaping gives a consistent lattice-surgery benefit in the budget sweep.} In the mixed-risk equal-budget sweep in Fig.~\ref{fig:e6-seam-cooling}, CAST is below both EM and SGB across the six plotted budgets, with EM/CAST ratios of $4.4$--$5.8\times$ and SGB/CAST ratios of $3.6$--$4.5\times$. The mechanism is targeted seam cooling, which relocates exposure away from merge-split observables under comparable total exposure. This supports seam-aware shaping for merge-split phases.

\subsection{Mechanism and Robustness Checks}

\textbf{CAST's components are complementary.} Table~\ref{tab:e7-mechanism-contribution} reports a component ablation on matched surface-code memory configurations, with pooled LER normalized to full CAST. In the table, Sens. denotes sensitivity weighting, Route denotes decoder-weighted-harm-aware route selection, Role denotes histogram-preserving rebinding of exposure to code roles, and Cool denotes boundary/seam cooling. The full configuration achieves the lowest LER ($(3.66\pm0.08)\times 10^{-3}$). Removing role rebinding causes the largest degradation ($2.20\times$ full CAST), followed by cooling controls ($1.68\times$) and route shaping ($1.44\times$). Sensitivity scoring alone or sensitivity plus routing remains near $1.9\times$ full CAST, showing that the harm model must be paired with placement actions.

\begin{table}[!t]
\centering
\caption{\textbf{Component ablation.} Pooled LER ($10^{-3}$) over nine matched surface-code memory configurations; errors are binomial standard errors and LER/Full is normalized to full CAST.}
\label{tab:e7-mechanism-contribution}
\scriptsize
\begin{tabular*}{\columnwidth}{@{\extracolsep{\fill}}cccc@{\hspace{4pt}\vrule width 0.35pt\hspace{4pt}}cc@{}}
\toprule
\multicolumn{4}{c@{\hspace{4pt}\vrule width 0.35pt\hspace{4pt}}}{Enabled CAST components} &
\multirow{2}{*}[-2pt]{LER $\pm$ SE ($10^{-3}$)} &
\multirow{2}{*}[-2pt]{LER/Full} \\
\cmidrule(r){1-4}
Sens. & Route & Role & Cool & & \\
\midrule
\textcolor{green!55!black}{\checkmark} & \textcolor{red!70!black}{$\times$} & \textcolor{red!70!black}{$\times$} & \textcolor{red!70!black}{$\times$} & 7.06 $\pm$ 0.14 & 1.93$\times$ \\
\textcolor{green!55!black}{\checkmark} & \textcolor{green!55!black}{\checkmark} & \textcolor{red!70!black}{$\times$} & \textcolor{red!70!black}{$\times$} & 6.98 $\pm$ 0.14 & 1.91$\times$ \\
\textcolor{red!70!black}{$\times$} & \textcolor{green!55!black}{\checkmark} & \textcolor{green!55!black}{\checkmark} & \textcolor{green!55!black}{\checkmark} & 4.70 $\pm$ 0.10 & 1.29$\times$ \\
\textcolor{green!55!black}{\checkmark} & \textcolor{red!70!black}{$\times$} & \textcolor{green!55!black}{\checkmark} & \textcolor{green!55!black}{\checkmark} & 5.28 $\pm$ 0.11 & 1.44$\times$ \\
\textcolor{green!55!black}{\checkmark} & \textcolor{green!55!black}{\checkmark} & \textcolor{red!70!black}{$\times$} & \textcolor{green!55!black}{\checkmark} & 8.04 $\pm$ 0.16 & 2.20$\times$ \\
\textcolor{green!55!black}{\checkmark} & \textcolor{green!55!black}{\checkmark} & \textcolor{green!55!black}{\checkmark} & \textcolor{red!70!black}{$\times$} & 6.14 $\pm$ 0.13 & 1.68$\times$ \\
\textcolor{green!55!black}{\checkmark} & \textcolor{green!55!black}{\checkmark} & \textcolor{green!55!black}{\checkmark} & \textcolor{green!55!black}{\checkmark} & \textbf{3.66 $\pm$ 0.08} & \textbf{1.00$\times$} \\
\bottomrule
\end{tabular*}
\end{table}

\begin{figure}[!t]
\centering
\includegraphics[width=\columnwidth]{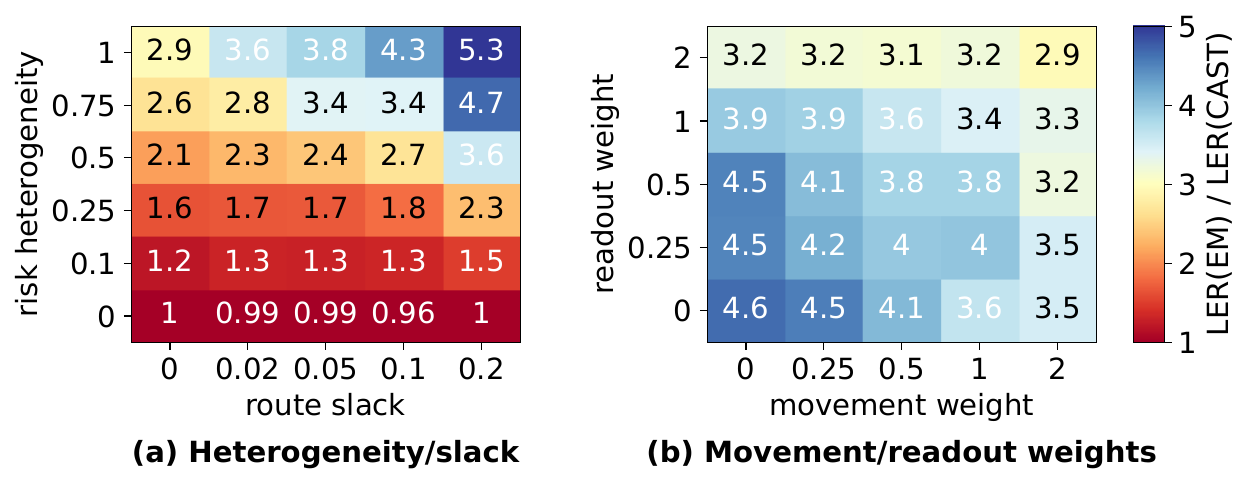}
\caption{\textbf{Hardware-sensitivity sweep.} EM/CAST LER ratios under (a) varying risk heterogeneity and route slack and (b) varying movement and readout loss weights. Values above one favor CAST.}
\label{fig:hardware-sensitivity}
\end{figure}

\textbf{CAST benefits most when loss risk is heterogeneous and routing has placement slack.} The hardware-sensitivity sweep in Fig.~\ref{fig:hardware-sensitivity} varies two conditions that determine whether deposition can be improved: risk heterogeneity and available routing slack. With zero heterogeneity, EM/CAST remains near one ($0.96$--$1.01\times$), indicating little placement headroom. As heterogeneity and slack increase, the ratio grows monotonically and reaches $5.3\times$. The movement/readout-weight sweep remains above one across the grid ($2.9$--$4.6\times$), showing that the benefit is not tied to a single physical loss source.

\begin{figure}[!t]
\centering
\includegraphics[width=0.8\columnwidth]{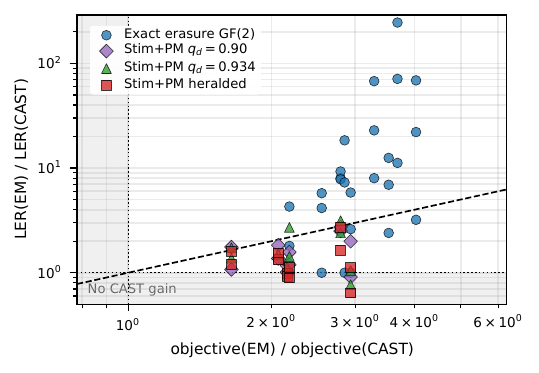}
\caption{\textbf{Decoder-mismatch validation.} EM/CAST ratios on identical traces: compiler objective on the $x$-axis and decoder LER on the $y$-axis.}
\label{fig:decoder-mismatch}
\end{figure}

\textbf{Decoder-weighted harm is a useful directional proxy for decoder-level gains.} The decoder-mismatch study in Fig.~\ref{fig:decoder-mismatch} compares EM/CAST ratios on identical traces, using the compiler objective on the $x$-axis and decoder LER on the $y$-axis. Most points lie above one on both axes; exact GF(2) erasure decoding shows the largest gains, while Stim+PyMatching remains mostly positive under imperfect and heralded detection. The spread across decoders shows that decoder-weighted harm guides placement, but final LER estimates still require decoder-level evaluation.

\section{Conclusion}
CAST frames atom loss as a placement problem: under comparable exposure, where loss lands can substantially affect logical failure. It combines trace exposure, decoder sensitivity, and local placement actions in a compiler-side pass. Across controlled memory, physical-scale, and lattice-surgery studies, CAST reduces logical failure, with decoder-weighted harm serving as a directional proxy for decoder-level performance. Future work will integrate CAST into full neutral-atom compiler backends and evaluate hardware-calibrated, compiler-generated workloads.

\bibliographystyle{IEEEtran}
\bibliography{reference}

\end{document}